\documentclass[cameraready]{Interspeech}
\usepackage{pifont}  
\usepackage{multirow}
\usepackage{booktabs}   
\usepackage{tabularx}   
\usepackage{makecell}   
\usepackage{array}     
\usepackage[ruled,vlined,linesnumbered]{algorithm2e}
\usepackage{amsmath}
\usepackage{amssymb} 
\SetAlgoSkip{0.6em}
\SetAlgoInsideSkip{0.4em}
\DontPrintSemicolon
\title{TASU2: Controllable CTC Simulation for Alignment and Low-Resource Adaptation of Speech LLMs}

\author[affiliation={1}, equalcontribution]{Jing}{Peng}
\author[affiliation={2}, equalcontribution]{Chenghao}{Wang}
\author[affiliation={1}]{Yi}{Yang}
\author[affiliation={1}]{Lirong}{Qian}
\author[affiliation={1}]{Junjie}{Li}
\author[affiliation={1}]{Yu}{Xi}
\author[affiliation={3}]{Shuai}{Wang}
\author[affiliation={1}, correspondingauthor]{Kai}{Yu}

\address{
  $^1$X-LANCE Lab, Department of Computer Science and Engineering, Shanghai Jiao Tong University, China\\
  $^1$MoE Key Lab of Artificial Intelligence,
  $^1$Jiangsu Key Lab of Language Computing\\
  $^2$AISpeech Ltd, Suzhou, China
  $^3$Nanjing University, China
}

\email{\{jing.peng, kai.yu\}@sjtu.edu.cn, voldbaboon@gmail.com, shuaiwang@nju.edu.cn}


\usepackage{comment}

\begin{document}

\maketitle

\begin{abstract}
Speech LLM post-training increasingly relies on efficient cross-modal alignment and robust low-resource adaptation, yet collecting large-scale audio-text pairs remains costly. Text-only alignment methods such as TASU reduce this burden by simulating CTC posteriors from transcripts, but they provide limited control over uncertainty and error rate, making curriculum design largely heuristic. We propose \textbf{TASU2}, a controllable CTC simulation framework that simulates CTC posterior distributions under a specified WER range, producing text-derived supervision that better matches the acoustic decoding interface. This enables principled post-training curricula that smoothly vary supervision difficulty without TTS. Across multiple source-to-target adaptation settings, TASU2 improves in-domain and out-of-domain recognition over TASU, and consistently outperforms strong baselines including text-only fine-tuning and TTS-based augmentation, while mitigating source-domain performance degradation.
\end{abstract}

\section{Introduction}
The rapid progress of large language models has accelerated Speech LLM research~\cite{Peng_2025,arora2025landscapespokenlanguagemodels}. 
However, strong Speech LLM performance often comes with heavy reliance on large-scale audio--text pairs and compute-intensive pipelines~\cite{radford2022robustspeechrecognitionlargescale}, making post-training, adaptation, and reproduction costly. 
Recent studies therefore revisit lightweight alignment between speech and text representations, including TASU~\cite{peng2026tasutextonlyalignmentspeech}, LegoSLM~\cite{ma2025legoslm}, and AlignFormer~\cite{fan2025alignformer}.

Among them, TASU (Text-only Alignment for Speech Understanding)  is particularly appealing because it enables \emph{text-only} post-training: it stochastically simulates CTC posteriors from transcripts, allowing training without paired audio while retaining real-audio inference. 
In practice, TASU can serve as an effective curriculum and improves recognition both on the source domain and under domain shift. 
Yet its simulation provides limited contro  l over posterior uncertainty and the resulting error regime, making difficulty scheduling largely heuristic.

In parallel, \emph{low-resource} speech understanding remains a persistent challenge~\cite{9801640}. 
Many target domains lack sufficient paired audio, and straightforward audio-based fine-tuning can yield unstable gains and noticeable \emph{source-domain degradation}~\cite{takashima2022updatingencoderspreventscatastrophic,burdisso2026textonlyadaptationllmbasedasr}. 
Text-centric adaptation has been explored to reduce the need for paired audio, including parameter-efficient tuning~\cite{liao2023zero} and text-only updates of the language component~\cite{fang2025low}. 
However, using plain text as the training signal still suffers from a mismatch to the acoustic decoding interface, and its improvements can be limited compared to stronger audio-based augmentation baselines such as TTS~\cite{casanova2023asrdataaugmentationlowresource}.

To address both limitations, we propose \textbf{TASU2}, a \emph{controllable} text-to-CTC simulation framework for Speech LLM post-training.
Instead of relying on unconstrained stochastic simulation, TASU2 generates pseudo CTC posterior distributions \emph{under a specified target WER range}, producing text-derived supervision that better matches the acoustic decoding interface.
This enables a principled curriculum that smoothly varies supervision difficulty and error profiles without TTS or paired audio.

Experiments show that TASU2 strengthens text-only alignment beyond TASU, improving recognition on the \emph{source domain} and under \emph{cross-domain generalization} without any audio-text training.
More importantly, under low-resource transfer settings, TASU2 consistently outperforms strong baselines including text-only adaptation~\cite{fang2025low} and TTS-based augmentation~\cite{casanova2023asrdataaugmentationlowresource}, while better preserving source-domain performance.

In summary, our contributions are:
\begin{itemize}
  \item We introduce a \emph{WER-conditioned} text-derived post-training signal by simulating calibrated CTC posterior distributions from transcripts, bridging the gap between plain text supervision and the acoustic decoding interface.
  \item We develop a controllable text-to-CTC simulator that generates posterior sequences under a specified WER range, enabling explicit control over supervision difficulty and error profiles for curriculum design.
  \item We demonstrate consistent gains over TASU on both source-domain and generalization evaluations without audio training, and competitive improvements in low-resource transfer over text-only and TTS-based augmentation baselines.
\end{itemize}

\begin{figure*}[!t]
  \centering
  \includegraphics[width=\textwidth,height=0.40\textheight,keepaspectratio]{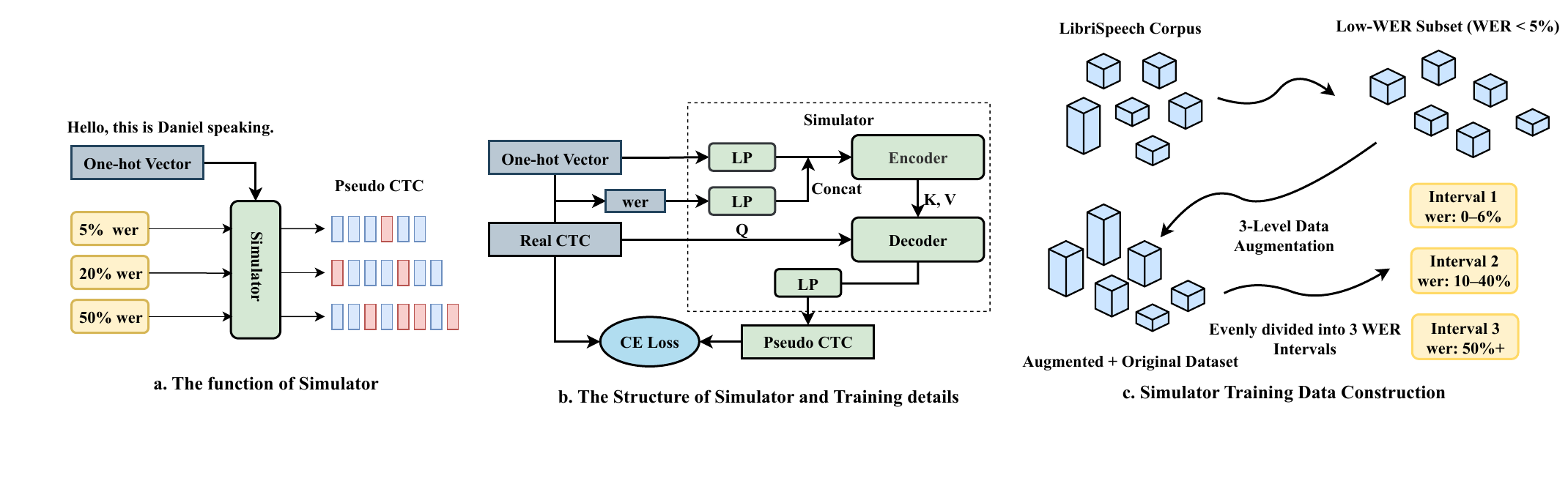}
  \caption{An Overview of TASU2.}
  \label{fig:structure_overview}
\end{figure*}

\section{Text-only Alignment: From TASU to TASU2}
\label{sec:related_work}
CTC introduces a blank symbol and marginalizes over alignments, collapsing frame-level posteriors into compact label sequences via blank removal and repetition merging~\cite{jung2023blankcollapsecompressingctc, deng2021improvinghybridctcattentionendtoend}. This compact representation has inspired several speech-LLM alignment methods, like AlignFormer and LegoSLM. However, these methods still rely on paired audio–text data and exhibit performance drops under limited supervision. 

TASU eliminates the need for paired data by simulating CTC posteriors from text alone. During inference, \emph{Label-Synchronous Decoding (LSD)} ~\cite{chen2016phone, deng2023labelsynchronousneuraltransduceradaptable} compresses real audio-derived CTC posteriors \(\mathbf{P}\in\mathbb{R}^{T\times V}\) by removing frames where the blank probability exceeds a threshold \(\tau\):
\begin{equation}
\mathbf{P}_t' = \begin{cases}\emptyset, & \text{if }P_t(<\text{blank}>)>\tau,\\ \mathbf{P}_t, &\text{otherwise},\end{cases}
\label{eq:lsd_removal}
\end{equation}
and then merges consecutive identical frames via averaging:
\begin{equation}
\mathbf{P}_t'' = \frac{1}{|S_j|}\sum_{t\in S_j}\mathbf{P}_t',\quad j=1,\dots,J.
\label{eq:lsd_merge}
\end{equation}
For text-only training, \emph{CTC Posterior Simulation (CPS)} generates pseudo-posteriors \(\tilde{\mathbf{S}}\) from token sequences by applying random label smoothing (\(\tilde{\mathbf{p}} = \alpha\delta_y + (1-\alpha)\frac{1}{V}\mathbf{1}\)), deletions, and insertions (blanks or duplicates). A projector trained on these simulated posteriors maps them into a frozen LLM, enabling zero-shot speech recognition and, when used as a curriculum pre-training stage, improved domain generalization.

Despite its effectiveness, TASU's simulation remains uncontrolled and may not fully capture real acoustic-phonetic confusions. This raises two questions: $controllability$ and $fidelity$, which we focus on in TASU2.
\section{TASU2: WER-Controllable Text-to-CTC Posterior Simulation}
\label{sec:method}

As motivated in related work, a key question in text-only alignment is whether a simulator can generate CTC-like posteriors that are \textbf{\emph{both} (i) close to real acoustic posteriors and (ii) controllable to support principled curricula and low-resource adaptation}. We address this with \textbf{TASU2}, which learns a \emph{controllable text-to-CTC posterior simulator} that outputs pseudo CTC posterior sequences conditioned on a transcript and a discrete WER control code. Fig.~\ref{fig:structure_overview} overviews the pipeline.

\subsection{Task and Notation}
\label{subsec:formulation}
Let the transcript be a token sequence $\mathbf{y}=(y_1,\ldots,y_U)$ from a tokenizer with vocabulary size $V$ (including the CTC blank). TASU2 learns a simulator that maps $(\mathbf{y},c)$ to a posterior-frame sequence $\hat{\mathbf{P}}=(\hat{\mathbf{p}}_1,\ldots,\hat{\mathbf{p}}_T)$, where $\hat{\mathbf{p}}_t\in[0,1]^V$ and $\sum_{v=1}^{V}\hat{\mathbf{p}}_{t,v}=1$. The control code $c\in\{1,\ldots,K\}$ indicates a target WER interval (e.g., 1,2,3 or low/medium/high).

\subsection{Training Signal: Distribution-level Supervision}
\label{subsec:objective}
As described in detail in Algorithm~\ref{alg:tasu2_sim}, for each utterance, a teacher ASR system provides a real CTC posterior sequence $\mathbf{P}=(\mathbf{p}_1,\ldots,\mathbf{p}_{T^\star})$. Since $T^\star$ varies, we pad to a fixed horizon $T_{\text{train}}$ with a validity mask $\mathbf{m}\in\{0,1\}^{T_{\text{train}}}$.
We train the simulator with posterior-level cross-entropy:
\begin{equation}
\mathcal{L}_{\text{sim}}(\theta)=
\mathbb{E}
\left[
-\frac{1}{\sum_t m_t}\sum_{t=1}^{T_{\text{train}}} m_t
\sum_{v=1}^{V} \mathbf{p}_{t,v}\log \hat{\mathbf{p}}_{t,v}
\right].
\label{eq:sim_ce}
\end{equation}
This \emph{distribution-matching} objective encourages CTC-like structure like blank dominance and token confusability, which is more faithful to acoustic decoding than text-only perturbations.

\subsection{Architecture Design of Simulator}
\label{subsec:arch}
We instantiate the simulator as a lightweight Transformer encoder--decoder (Fig.~\ref{fig:structure_overview}(b)). The transcript is embedded and combined with a learned embedding for the WER code $c$, which conditions generation. The decoder autoregressively outputs one posterior frame per step. During training we use teacher forcing; at inference time, the simulator runs autoregressively conditioned only on $(\mathbf{y},c)$.

\subsection{WER-conditioned Data Construction}
\label{subsec:data}
To obtain controllable supervision, we need the \emph{same} transcript paired with teacher posteriors spanning different error regimes. Following Fig.~\ref{fig:structure_overview}(c), we sample only \emph{a one-seventh} portion of LibriSpeech and generate multiple augmented variants (noise/reverb). For each variant, we run a teacher ASR model to obtain a greedy hypothesis $\tilde{\mathbf{y}}$ and CTC posteriors $\mathbf{P}$. We compute WER$(\tilde{\mathbf{y}},\mathbf{y})$ and map it to one of $K$ WER intervals to form the control code $c$. Using discrete intervals reduces sensitivity to measurement noise and alleviates skewed WER distributions, yielding a more stable control signal. And we train the simulator on this data.

\begin{algorithm}[t]
\caption{TASU2: WER-Conditioned Text-to-CTC Simulation}
\label{alg:tasu2_sim}
\small
\KwIn{Transcript $\mathbf{y}$, vocab size $V$ (incl. blank), WER bins $\{\mathcal{I}_k\}_{k=1}^K$, teacher ASR $\mathcal{T}$, augmentor $\mathcal{A}$, horizon $T_{\text{train}}$}
\KwOut{Simulator parameters $\theta$}
\BlankLine
\textbf{(1) Build multi-WER supervision}\;
\ForEach{utterance $(x,\mathbf{y})$}{
  Generate augmented waveforms $\{x^{(j)}\}\leftarrow \mathcal{A}(x)$\;
  \ForEach{$x^{(j)}$}{
    Obtain teacher posteriors $\mathbf{P}^{(j)} \leftarrow \mathcal{T}_{\text{CTC}}(x^{(j)})$\;
    Obtain greedy hypothesis $\tilde{\mathbf{y}}^{(j)} \leftarrow \mathcal{T}_{\text{greedy}}(x^{(j)})$\;
    Compute $\mathrm{WER}^{(j)}=\mathrm{WER}(\tilde{\mathbf{y}}^{(j)},\mathbf{y})$ and assign $c^{(j)}\leftarrow \mathrm{bin}(\mathrm{WER}^{(j)};\{\mathcal{I}_k\})$\;
    Store tuple $(\mathbf{y},c^{(j)},\mathbf{P}^{(j)})$\;
  }
}
\BlankLine
\textbf{(2) Train conditional simulator}\;
\While{not converged}{
  Sample a tuple $(\mathbf{y},c,\mathbf{P})$\;
  Pad $\mathbf{P}$ to length $T_{\text{train}}$ and build mask $\mathbf{m}$\;
  Predict $\hat{\mathbf{P}} \leftarrow p_\theta(\cdot \mid \mathbf{y},c)$ with teacher forcing\;
  Update $\theta$ by minimizing $\mathcal{L}_{\text{sim}}$ in Eq.~\eqref{eq:sim_ce}\;
}
\end{algorithm}

\subsection{Simulator Fidelity Analysis}
\label{subsec:fidelity}
We further assess whether the simulator indeed reproduces \emph{acoustic-like} CTC posteriors and whether the WER control behaves as intended. We consider two complementary aspects:

\noindent\textbf{(i) Posterior similarity to teacher CTC (unconditioned).}
We compare TASU2 against the TASU baseline simulation (without WER conditioning) using distribution-level metrics between teacher posteriors $\mathbf{P}$ and simulated posteriors $\hat{\mathbf{P}}$:

\begin{equation}
\begin{aligned}
\mathrm{CE} &= \frac{1}{N}\sum_{t=1}^{N}\Bigl(-\sum_{v=1}^{V} p_{t,v}\log \hat{p}_{t,v}\Bigr), \\
\mathrm{KL} &= \frac{1}{N}\sum_{t=1}^{N}\sum_{v=1}^{V} p_{t,v}\log \frac{p_{t,v}}{\hat{p}_{t,v}}, \\
\mathrm{Acc} &= \frac{1}{N}\sum_{t=1}^{N}\mathbb{I}\bigl[\arg\max_{v} p_{t,v}=\arg\max_{v} \hat{p}_{t,v}\bigr], \\
\mathrm{ProbDiff} &= \frac{1}{N}\sum_{t=1}^{N}\bigl|\max_{v} p_{t,v}-\max_{v} \hat{p}_{t,v}\bigr|,
\end{aligned}
\end{equation}
where $N$ is the number of valid frames. As shown in Table~\ref{tab:simulator_fidelity}, TASU2 yields substantially lower CE/KL and improved argmax agreement, indicating closer matching to real CTC posteriors.

\begin{figure}[t]
  \centering
  \includegraphics[width=\linewidth]{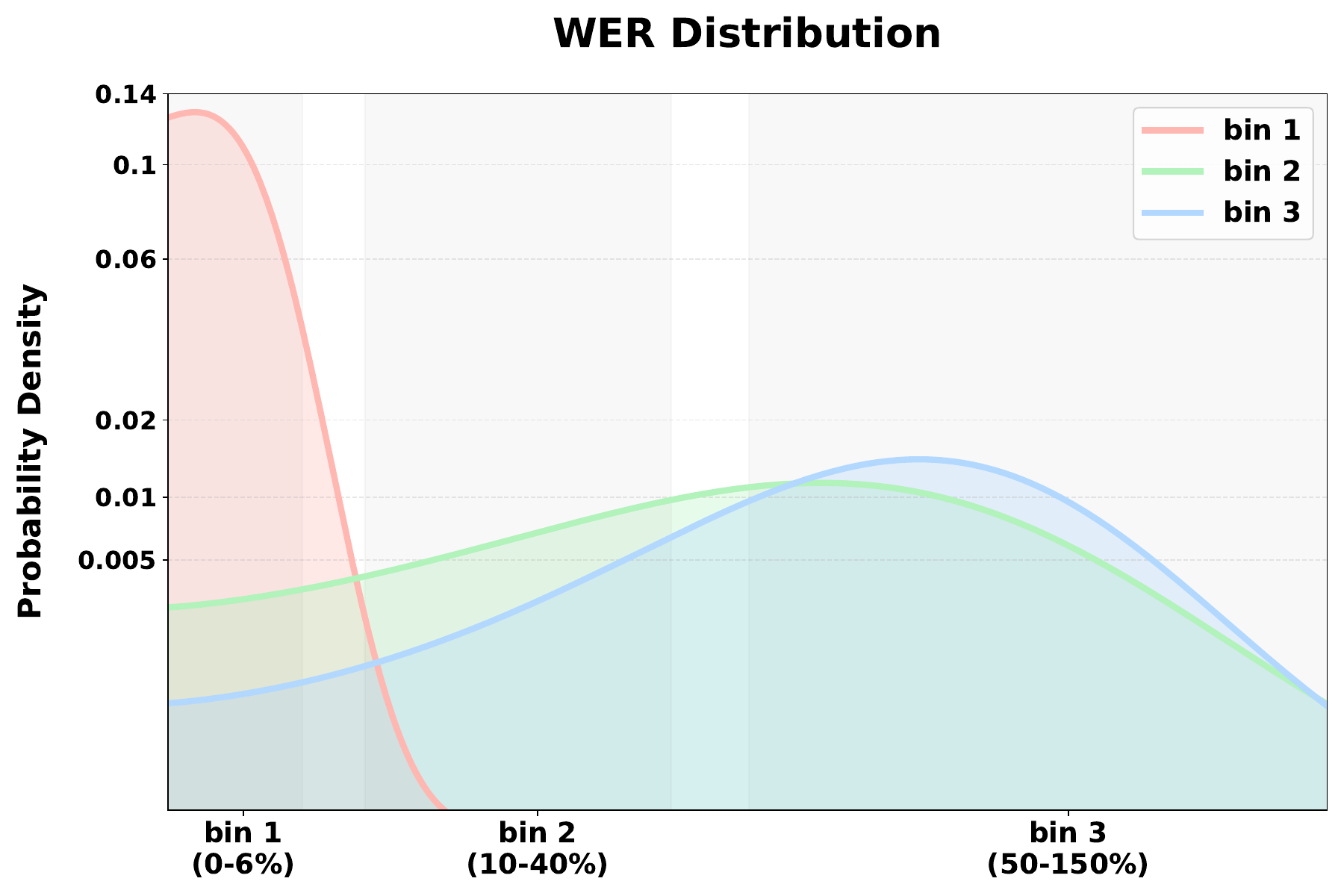}
  \caption{WER controllability under WER-bin conditioning.}
  \label{fig:wer_bin}
\end{figure}

\noindent\textbf{(ii) WER controllability (conditioned).}
To verify controllability, we inject WER-bin codes and decode the simulated posteriors with a fixed decoder, then measure the realized WER for each bin. Fig.~\ref{fig:wer_bin} reports the empirical WER distributions across bins, showing that TASU2 reliably separates error regimes and tracks the target intervals.


\begin{table}[t]
  \centering
  \caption{Posterior similarity metrics (lower is better except Acc).}
  \label{tab:simulator_fidelity}
  \small
  \resizebox{\linewidth}{!}{%
  \begin{tabular}{lcccc}
    \toprule
    Method & CE $\downarrow$ & KL $\downarrow$ & Acc $\uparrow$ & ProbDiff $\downarrow$ \\
    \midrule
    TASU-style baseline & 2.5149 & 2.3372 & 0.8220 & 0.0813 \\
    TASU2 (AR simulator) & \textbf{1.2296} & \textbf{1.0497} & \textbf{0.8782} & \textbf{0.0615} \\
    \bottomrule
  \end{tabular}%
  }
\end{table}
\section{Experiments}
To validate the effectiveness and controllability of TASU2, we evaluate (i) text-to-CTC alignment quality and cross-domain generalization, and (ii) low-resource domain adaptation.

\subsection{Model Architecture}
Our simulator is a Transformer encoder--decoder with 6 encoder and 6 decoder layers and hidden size 512. It takes a transcript and a WER-bin ID as input and autoregressively generates pseudo CTC posterior frames. For the Speech LLM, we follow TASU~\cite{peng2026tasutextonlyalignmentspeech} and use SenseVoice-Small~\cite{an2024funaudiollmvoiceunderstandinggeneration} as the speech encoder and Qwen2.5-1.5B~\cite{qwen2025qwen25technicalreport} as the LLM, connected by a Linear--SiLU--Linear projector.

\subsection{Datasets}
We pre-train the CTC simulator on one-seventh of LibriSpeech (960h)~\cite{7178964} with augmentation, to learn a general mapping from linguistic units to acoustic-like posterior distributions. For Stage~2 post-training, we synthesize pseudo CTC supervision from text-only corpora in LibriSpeech and Medical~\cite{mooney2019medical}. We evaluate on LibriSpeech (test-clean/other), Medical (8h), TED-LIUM 3~\cite{Hernandez_2018}, SlideSpeech~\cite{wang2023slidespeechlargescaleslideenrichedaudiovisual}, and CoVoST2 En$\rightarrow$Zh~\cite{wang2021covost2}.

\subsection{Training and Evaluation Setup}
\noindent\textbf{WER-conditioned simulation.}
We discretize the realized teacher WER into three bins: 0--6\% (low), 10--40\% (medium), and 50--150\% (high). Unless otherwise stated, we use bin~1 for conditioning in the main experiments.

\noindent\textbf{Two-stage post-training.}
Stage~1 trains on simulated supervision with learning rate $5\times 10^{-5}$ for 5 epochs. Stage~2 adapts the model using simulated text from the target domain (Medical or SlideSpeech), following the same optimization protocol.

\noindent\textbf{Optimization.}
We use AdamW with DeepSpeed ZeRO-2 on 8$\times$Ascend 910B NPUs. LoRA (rank=16, $\alpha$=32) is used only for the two-stage adaptation setting in Table~\ref{tab:exp2_two_stage}.
\section{Evaluation And Analysis}

We evaluate TASU2 from three perspectives. Table~\ref{tab:wer_main} probes whether simulated posteriors improve CTC-style alignment and cross-domain generalization without audio training. Table~\ref{tab:multitask_tasu2} provides a lightweight multi-task sanity check beyond ASR. Our key result is Table~\ref{tab:exp2_two_stage}, where TASU2 achieves strong low-resource target gains while largely preserving source-domain performance, outperforming both text-only adaptation and TTS-based augmentation.

\begin{table}[t]
\centering
\caption{\textbf{Alignment and generalization of simulated CTC training.}
All models are trained on LibriSpeech and evaluated on LibriSpeech, SlideSpeech, and TED-LIUM; WER (\%)$\downarrow$ is reported.
\textbf{TASU2 conditions}: \textsuperscript{\ding{108}}~None (no WER conditioning);
\textsuperscript{\ding{109}}~WER-binned (coarse 3-level WER bins; we use bin~1 here).}
\label{tab:wer_main}

\setlength{\tabcolsep}{3pt}
\renewcommand{\arraystretch}{1.1}

\resizebox{\columnwidth}{!}{%
\begin{tabular}{l cc c c c}
\toprule
\multirow{2}{*}{\textbf{System}} &
\multicolumn{2}{c}{\textbf{Train Data}} &
\textbf{Libri} &
\multirow{2}{*}{\textbf{Slide}} &
\multirow{2}{*}{\textbf{TED}} \\
& \textbf{Text} & \textbf{(Audio, Text)} & \textit{clean/other} & & \\
\midrule
TASU
& Libri & --   & 4.57 / 9.90          & 24.07         & 19.36 \\
TASU
& Libri+Slide & --   & 4.21 / 10.31 & 18.70 & \textbf{13.23} \\
SLAM-CTC
& --   & Libri & \textbf{3.13} / 8.59 & 18.59         & 14.61 \\
\midrule
\textbf{TASU2}\textsuperscript{\ding{108}}
& Libri & --   & 4.63 / 9.82          & \textbf{16.41} & \textbf{14.02} \\
\textbf{TASU2}\textsuperscript{\ding{109}}
& Libri & --   & 3.41 / \textbf{8.15} & 17.67          & 14.49 \\
\textbf{TASU2}\textsuperscript{\ding{109}}
& Libri+Slide & --   & 3.94 / 8.25 & 17.31         & 13.93 \\
\bottomrule
\end{tabular}%
}
\end{table}

\subsection{Alignment and Generalization}
\label{sec:alignment}
We evaluate simulated CTC training on both in-domain and out-of-domain test sets to probe \emph{alignment} and \emph{generalization} of TASU2. We start from ASR and further extend the evaluation to a multi-task speech understanding setting. 

\noindent\textbf{ASR performance.}
As illustrated in Table~\ref{tab:wer_main}, relative to the original TASU random simulation, TASU2 consistently improves in-domain recognition while yielding much stronger cross-domain transfer. In particular, TASU2 reduces WER markedly on \emph{SlideSpeech} and \emph{TED-LIUM3}, and in several cases becomes competitive with, or even surpasses, the strong SLAM-CTC audio baseline, despite using text-only training inputs. 
Moreover, when we expand the text side by adding SlideSpeech transcripts (Libri+Slide), TASU2 further improves on the target and TED sets, indicating that text expansion and posterior simulation are complementary: additional target-style text benefits transfer, while TASU2 provides the CTC-like supervision signal needed to make such text effective.

To understand the role of controllability, we compare two simulator variants for generating pseudo posteriors: \emph{unconditioned} simulation (None) and \emph{WER-binned} simulation (coarse 3-level bins; we use bin~1 here). 
The ablation reveals a clear trade-off. The unconditioned simulator tends to improve out-of-domain recognition (better Slide/TED WER), but it incurs noticeable regression on the source domain (worse Libri). 
In contrast, WER-binned conditioning yields a better balance between alignment and transfer: it preserves strong generalization while substantially reducing source-domain degradation. 
These results support our design choice that lightweight, discrete error control is useful not only for curriculum design, but also for stabilizing domain shift behavior in simulated-CTC training.

\begin{table}[htb]
  \centering
  \caption{\textbf{TASU2 performance under multi-task evaluation.}
  \textbf{TASU2} uses WER-conditioned simulation with discretized WER bins; here we use only \textbf{bin~1} for conditioning.}
  \label{tab:multitask_tasu2}
  \setlength{\tabcolsep}{5pt}
  \renewcommand{\arraystretch}{1.12}
  \resizebox{\columnwidth}{!}{%
  \begin{tabular}{lccc}
    \toprule
    \textbf{Model} &
    \shortstack[c]{\textbf{Train audio}\\\textbf{duration(h)}} &
    \shortstack[c]{\textbf{LibriSpeech} $\downarrow$\\ \scriptsize clean/other (WER$\downarrow$)} &
    \shortstack[c]{\textbf{CoVoST2} $\uparrow$\\ \scriptsize En2Zh (BLEU$\uparrow$)} \\
    \midrule
    TASU         & 0       & 6.47 / 10.35 & 33.35 \\
    TASU2        & 0       & 3.68 / 8.25  & 33.08 \\
    SLAM         & 1.8k    & 3.30 / 7.24  & 37.34 \\
    Step-Audio   & $>$1000k & 2.36 / 6.32  & -- \\
    Qwen2.5-Omni & $>$1000k & 2.37 / 4.21  & 41.40 \\
    \bottomrule
  \end{tabular}%
  }
\end{table}

\noindent\textbf{Multi-task performance.}
Table~\ref{tab:multitask_tasu2} provides a lightweight multi-task sanity check. Since TASU2 simulates CTC posteriors with WER-conditioned control, its supervision is most directly aligned with token recognition and decoding, so gains beyond ASR are not necessarily expected. Still, TASU2 substantially improves over TASU on LibriSpeech without using any training audio, and remains competitive on CoVoST2, suggesting that CTC-style simulated supervision can transfer modestly beyond pure ASR under a zero-audio regime.

\newcolumntype{Y}{>{\centering\arraybackslash}X}  
\newcolumntype{C}[1]{>{\centering\arraybackslash}p{#1}} 

\begin{table}[t]
\centering
\caption{\textbf{Two-stage domain adaptation (source $\rightarrow$ target).}
All systems are first trained on the source domain (LibriSpeech) and then adapted to the target domain (Medical). We report WER\%$\downarrow$. LoRA is used for all systems for better performance.}
\label{tab:exp2_two_stage}
\setlength{\tabcolsep}{4pt}
\renewcommand{\arraystretch}{1.08}
\small

\resizebox{\columnwidth}{!}{%
\begin{tabular}{l c c c c}
\toprule
\textbf{System} & \multicolumn{2}{c}{\textbf{Train data}} & \textbf{LibriSpeech} & \textbf{Medical} \\
& \textbf{Stage 1} & \textbf{Stage 2} & \textit{clean/other} & \textit{test} \\
\midrule

\multirow{4}{*}{\textsc{SLAM-CTC}}
& \multirow{4}{*}{Audio} & --                          & 2.43 / 6.07 & 17.55 \\
&                        & Raw text~\cite{fang2025low}  & 2.56 / 6.62 & 13.62 \\
&                        & TTS audio                    & 2.72 / 6.80 & 12.79 \\
&                        & Raw audio                    & 2.70 / 6.76 & 12.35 \\
\midrule

\multirow{2}{*}{\textbf{\textsc{TASU2\textsuperscript{\ding{109}}}}}
& Text Sim-CTC & --           & 2.94 / 7.16 & 15.34 \\
& Text Sim-CTC & Text Sim-CTC  & 2.96 / 7.23 & \textbf{12.12} \\
\bottomrule
\end{tabular}%
}
\end{table}

\subsection{Domain Adaptation}
Section~\ref{sec:alignment} indicates that TASU2 achieves stronger alignment and generalization. In this section, we turn to its domain adaptation capability, with a particular focus on low-resource settings. We study two-stage domain adaptation from a source domain to a low-resource target domain.
LibriSpeech serves as the source-domain pre-training data (Stage~1), while the medical set simulates a low-resource target domain for adaptation (Stage~2).
We report WER on both LibriSpeech and the medical test set to jointly measure \emph{target improvement} and \emph{source retention}.

Table~\ref{tab:exp2_two_stage} highlights a clear trade-off in conventional adaptation: target-domain gains often come with reduced source retention. For \textsc{SLAM-CTC}, adapting with target-domain audio (raw or TTS) improves the Medical WER to 12.35/12.79, but also shifts LibriSpeech from 2.43/6.07 to around 2.70--2.72/6.76--6.80, indicating a non-trivial source-domain drop.

In contrast, TASU2 yields a more favorable low-resource transfer profile. Using only text-derived simulated CTC supervision for Stage~2, TASU2 achieves the best Medical WER of \textbf{12.12}, outperforming text-only adaptation (13.62) and TTS augmentation (12.79), and even slightly surpassing raw-audio adaptation (12.35). Meanwhile, source-domain WER remains nearly unchanged from 2.94/7.16 to 2.96/7.23 (only +0.02/+0.07), demonstrating strong source retention. These results suggest that WER-controlled posterior simulation provides an effective and stable alternative to target-audio-heavy fine-tuning when paired data is scarce.

\section{Conclusion}
\label{sec:conclusion}

We presented \textbf{TASU2}, a WER-controllable text-to-CTC simulator trained with distribution-level supervision. By generating pseudo posteriors that better match acoustic CTC behavior, TASU2 provides stronger alignment signals for speech foundation models and Speech LLMs. Across evaluations, it improves robustness and cross-domain generalization, and enables effective source-to-target adaptation. In low-resource targets, TASU2 can outperform TTS-based augmentation, offering a practical alternative when paired audio is scarce.

\newpage
\section{Generative AI Use Disclosure}

During the preparation of this work, we used generative AI tools for assistance. The AI tools were only employed for improving the presentation, readability, and formatting of the manuscript, as well as for auxiliary support in code development and verification. They were not used to generate any substantial content, core ideas, experimental design, analysis, or conclusions of the paper.
\bibliographystyle{IEEEtran}
\bibliography{mybib}

\end{document}